\documentclass[pdflatex,sn-nature]{sn-jnl}
\usepackage{graphicx}
\usepackage{subcaption}
\usepackage{float}
\usepackage{rotating}
\usepackage{multirow}
\usepackage{amsmath,amssymb,amsfonts}
\usepackage{bbm}
\usepackage{amsthm}
\usepackage{mathrsfs}
\usepackage[title]{appendix}
\usepackage{xcolor}
\usepackage{textcomp}
\usepackage{manyfoot}
\usepackage{booktabs}
\usepackage{algorithm}
\usepackage{algorithmicx}
\usepackage{algpseudocode}
\usepackage{listings}
\theoremstyle{thmstyleone}%

\theoremstyle{thmstyletwo}%
\theoremstyle{thmstylethree}%
\begin{document}
\title[Bidirectional BTE--Fourier Coupling via Neural Operators]{
Neural-operator acceleration of bidirectional phonon Boltzmann–Fourier coupling for multiscale heat transport in silicon fin arrays}

\author[1]{\fnm{Ruthwik} \sur{Chivukula}}

\author[2]{\fnm{Khemraj} \sur{Shukla}}

\author*[2]{\fnm{George Em} \sur{Karniadakis}}\email{george\_karniadakis@brown.edu}

\affil[1]{\orgdiv{School of Engineering}, \orgname{Brown University}, \orgaddress{\street{184
Hope Street}, \city{Providence}, \postcode{02912}, \state{RI}, \country{USA}}}

\affil*[2]{\orgdiv{Division of Applied Mathematics}, \orgname{Brown University},
\orgaddress{\street{182 George Street}, \city{Providence}, \postcode{02912}, \state{RI},
\country{USA}}}

\abstract{
Localized heating in semiconductor devices can produce non-diffusive phonon transport near heat-generating regions, while heat spreading through the surrounding substrate remains well described by Fourier conduction. We present a hybrid framework that couples a pretrained neural operator for the phonon Boltzmann transport equation (BTE) within individual fins to a Fourier model of a shared substrate. To reduce the cost of repeated BTE solves across fin positions and coupling iterations, a DeepONet maps each fin’s substrate-interface temperature and heating state to its temperature and heat-flux fields. The operator is trained on interface temperatures generated by a linearized proxy of the coupled problem, which avoids explicit training data collection from full coupled runs. An energy-balance penalty regularizes the learned operator, and a mixed finite-element formulation represents substrate heat flux in an H(div)-conforming space. The two models exchange temperature and normal heat flux through a non-overlapping Schwarz iteration with adaptive Aitken relaxation.
We evaluate the framework on an idealized three-dimensional array of 100 silicon fins. For a held-out switching configuration, the neural-operator–Fourier solution differs from the direct BTE–Fourier reference by a maximum of 0.3 K in temperature and 0.6\% in relative flux across all fins. Excluding the one-time substrate factorization, the coupled surrogate run takes about 50 s, compared with 2 h 28 min for direct coupling, a speedup of approximately 180\texttimes{}. Additional held-out switching patterns converge with the same trained operator. These results demonstrate neural-operator acceleration of bidirectional kinetic–continuum coupling for the tested geometry and material.
This bidirectional strategy may extend to other multiscale materials problems coupling learned local neural models with continuum solvers.
}

\maketitle

\section{Introduction}\label{sec1}

The continuing rise in transistor density has been accompanied by a marked
reduction in device dimensions. Self-heating due to Joule dissipation is
consequently concentrated within increasingly small regions, whose characteristic
length scales may approach the mean free paths of the phonons responsible for heat transport. In the vicinity of these localized hot spots, heat transport departs from the diffusive limit, becoming ballistic or quasi-ballistic as boundary and non-equilibrium effects become appreciable, rendering the Fourier conduction law unfit for thermal modeling \cite{Moore1965,Schaller1997,Ahn2018,Swahn2008}. The resulting discrepancy is of practical concern as the thermal state of the device influences electrical characteristics and power dissipation behavior \cite{Raskin2014SOIMOSFET,Yin2018FinFET}. Non-uniform temperature fields and differences in thermal expansion further
give rise to thermo-mechanical stresses within the device and its surrounding
structure \cite{Burenkov2016Thermomechanical}. Accurate prediction of
temperature and heat flux is therefore important not only for thermal
management, but also for determining safe operating conditions and assessing
the long-term reliability of semiconductor and power-electronic devices
\cite{Ciappa2002Failure}.

The non-diffusive character of heat transport is, however, local. The small
active transistor regions are supported by a much larger substrate, and sufficiently far
from the heat source the effects of finite phonon mean free paths subside and
transport returns to the diffusive regime
\cite{Sverdrup2000,Pop2001,Donmezer2014,DonmezerGraham2014}. The phonon
Boltzmann transport equation (BTE), which describes the non-equilibrium phonon
population, accounts for ballistic and quasi-ballistic transport and recovers
the diffusive limit at larger length scales \cite{Peraud2011,Allu2018}. It may
therefore be used throughout the structure in principle, which would however be
computationally wasteful, since deep within the substrate, where the
phonon distribution has relaxed towards local equilibrium, the additional
kinetic resolution of the BTE offers little advantage over Fourier conduction
\cite{Donmezer2014,Swahn2008}.

This separation of transport regimes suggests a heterogeneous description, where the BTE may be confined to those regions in which non-diffusive effects remain appreciable, while Fourier conduction is retained elsewhere. The boundary separating these descriptions is artificial, and its location is itself a modeling choice. Hybrid BTE--Fourier descriptions have previously been considered through
several forms of decomposition, including spatial partitions about local hot spots and mode-space partitions based on phonon mean free path or Knudsen number \cite{DonmezerGraham2014,Vallabhaneni2017,Loy2013}. In a spatial decomposition, consistency between the two descriptions requires suitable transmission conditions at the artificial interface, where both temperature and normal heat flux must agree. However, a temperature jump arises at interfaces between different materials, as a consequence of thermal boundary (Kapitza) resistance \cite{Swartz1989}. The transmission conditions may be established iteratively by passing boundary data between the subdomains until the interface quantities cease to change. 
More generally, this alternating exchange belongs to the class of Schwarz domain-decomposition methods \cite{Lions1988SchwarzI,Gander2008}, which have a long history in numerical analysis and in the coupling of heterogeneous numerical models. These strategies establish multiscale coupling frameworks not limited to heat transfer alone, but can be extended to several engineering applications.

Up to now, the majority of the existing literature has considered the coupling of a small number of kinetic and continuum subdomains. In a multi-fin system, however, the substrate is shared by numerous spatially separated kinetic regions, each requiring a BTE solution at every coupling iteration. The computational cost consequently grows with the number of fins as well as the number of iterations. For fins of identical geometry and material, these local BTE problems differ only in the boundary temperature prescribed by the substrate, for a given heating state. Each calculation may therefore be viewed as an evaluation of the same solution operator,
$\mathcal{B}\colon \left(\left.T\right|_{\Gamma}, \mathbbm{1}_{\{\dot{q}=\dot{q}_{\mathrm{on}}\}}\right)\longmapsto (T,\mathbf{J})$ which maps the prescribed interface temperature and heat source to the temperature and heat-flux fields within the fin. This construction is closely related to the classical Dirichlet-to-Neumann map encountered in domain decomposition \cite{Toselli2005}, although the operator considered here returns the interior fields in addition to the boundary flux. The expense of repeated BTE calculations may thus be approached as a problem of approximating this local solution operator over the range of boundary traces encountered during the coupled calculation.

Neural operators provide a means of constructing such an approximation by learning mappings between function spaces from representative input-output pairs \cite{Lu2021,Li2021,Kovachki2023}. For the present problem, a single neural operator may be trained over a family of interface-temperature traces and heating states, and subsequently reused across fins as their boundary conditions evolve during the coupling iteration. The availability of public phonon-transport solvers, including OpenBTE and GiftBTE, provides a practical means of generating the BTE solutions required for supervised training \cite{Romano2021,Hu2023}. While physics-informed approaches to solving the phonon BTE have also been investigated \cite{Li2022PINNBTE,Zhou2023,Li2023}, we retain supervised operator learning from BTE solutions, with energy conservation imposed through an additional regularization term.

Learned neural surrogates have increasingly been used in conjunction with conventional
numerical solvers, coupling trainable and conventional numerical methods either through
temporal alternation \cite{Oommen2022,Oommen2024} or through a spatial
decomposition \cite{Yin2022,Wang2025,Wang2026,Yang2026DDDeepONet,Shukla2025}. Pretrained DeepONets have likewise been employed to 
replace a large number of finite elements (FEs) with far fewer neural operator elements (NOEs), accelerating the joint energy functional minimization problem by reducing the degrees of freedom \cite{Ouyang2026NOEM}. 
These studies establish the broader possibility of interweaving a learned solution operator into a numerical framework.

In most of the works above, the interacting domains share the same governing physics. Here, we consider a setting where the subdomains obey different transport descriptions, and the learned kinetic operator is invoked repeatedly over many spatially separated local subdomains which communicate through a shared continuum domain. A closely related work is that of Panda \textit{et al.}~\cite{Panda2026}, who considered a data-driven DeepONet for nanoscale FinFET thermal analysis. Their construction exploits the repeated structure of a fin array, establishing the possibility of reusing an operator across
different fin locations and thermal states. However, their work uses only a one-way information flow from substrate to the fin, leaving the scope of bidirectional coupling unaddressed. The question considered here is therefore whether a reusable local neural operator can be placed within a two-way kinetic--continuum data transmission. In such a framework, errors in exchanged quantities may accumulate or
disturb convergence, especially with several subdomains interacting with a single large domain, which makes this a challenging yet interesting research topic to pursue.

We consider these questions on an idealized micron-scale three-dimensional array 
of 100 fins sharing a common substrate in which non-diffusive transport remains 
significant within the fins but gives way to diffusion in the substrate. This allows 
the kinetic--continuum interface to coincide with the fin--substrate boundary, while 
retaining the distinct transport physics that necessitate a heterogeneous description. 
Each fin is handled by phonon BTE, while Fourier conduction in the substrate is discretized 
using a mixed finite-element formulation. The repeated fin-scale BTE solution is replaced by 
a single neural operator, which is queried independently at each fin during the Schwarz
iteration. We refer to the resulting framework as NO--Fourier (neural operator--Fourier), in contrast to direct BTE--Fourier coupling, and use a DeepONet as the neural operator. To avoid collecting training data from full coupled runs, the interface temperatures used for training are generated from a linearized estimate of the interface response over many switching patterns. Heat flux in the substrate is represented in an
\(H(\mathrm{div})\)-conforming space, so that its normal component may be
imposed as an essential condition at the artificial interfaces. The operator is trained with an
additional control-volume energy-conservation term, which behaves as a physics-informed regularization term \cite{Raissi2019}. The overall framework
therefore promotes conservation in the local surrogate through the energy penalty, and recovers flux continuity at convergence in the exchange
between the kinetic and continuum descriptions. Comparisons are made against
direct BTE--Fourier coupling on the same 100-fin problem, considering field
errors, interface temperature mismatch, normal-flux continuity, energy balance,
convergence of the Schwarz iteration and computational cost.

The scope of this approach extends beyond phonon transport. Similar
multiscale settings arise in materials mechanics, where atomistic processes
near crack tips or defect cores interact with a larger region described by
continuum elasticity \cite{Tadmor1996,Broughton1999}. In such settings,
learning the local material response offers a way to reduce the cost of
repeated fine-scale solves while retaining their bidirectional coupling
with the surrounding continuum.

\section{Results}\label{sec:results}

\subsection{Problem setup}\label{subsec:problem_setup}

\begin{figure}
    \centering
    \includegraphics[width=\linewidth]{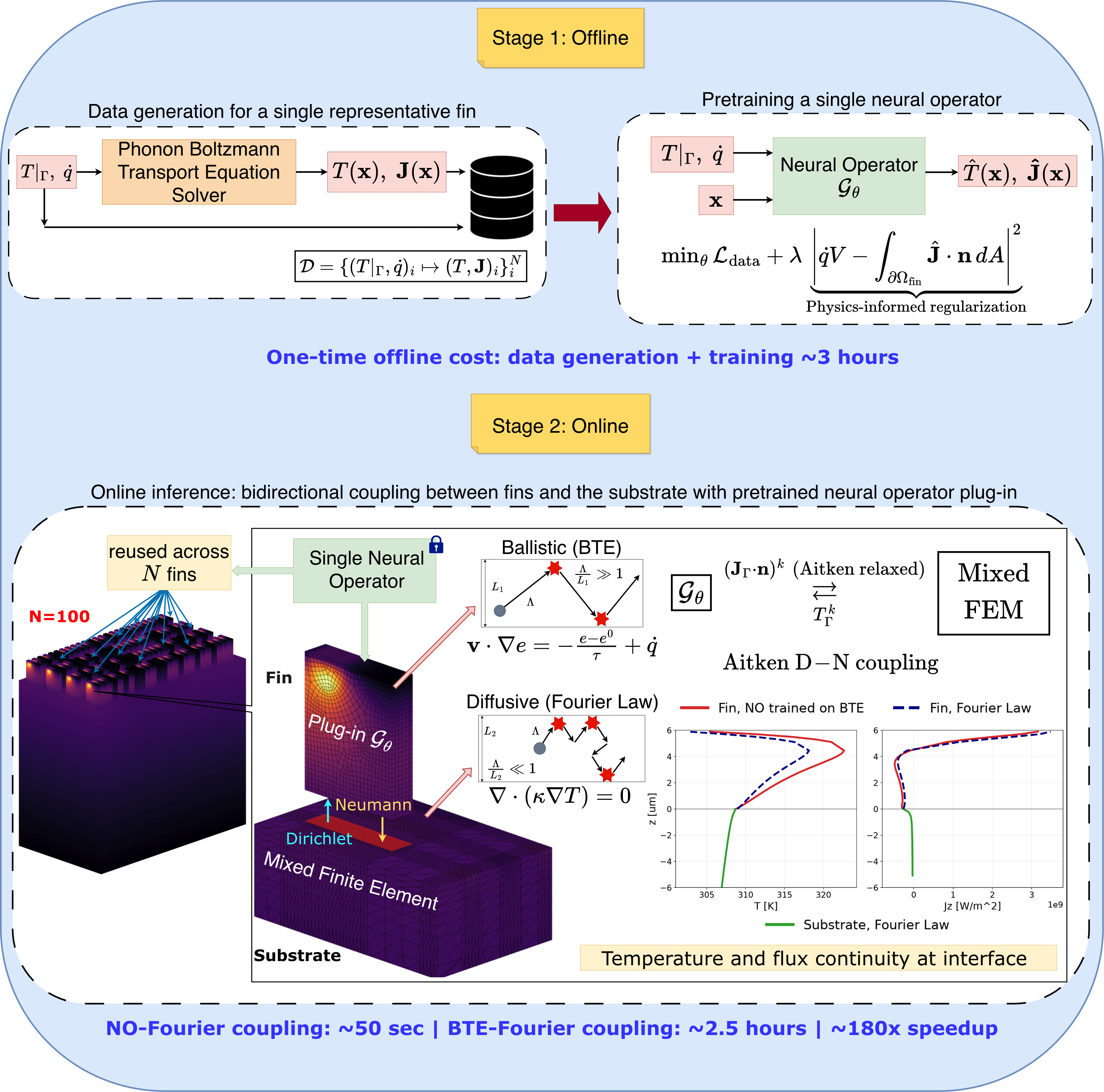}
    \caption{Overview of the framework. Offline GiftBTE solves representative fin problems to generate training data for a neural operator with a DeepONet backbone. Online, the pretrained operator is reused across all $N=100$ fins and coupled to the mixed-FEM substrate through a non-overlapping Dirichlet--Neumann Schwarz iteration with adaptive Aitken relaxation. Coupling time reported above excludes the one-time LU factorization cost.}
    \label{fig:framework}
\end{figure}

We consider a representative three-dimensional steady-state thermal problem
consisting of a $4\times25$ array of 100 fin field-effect transistors (FinFETs) distributed over a
common silicon substrate, as shown in Fig.~\ref{fig:framework}.
Each fin represents the active region associated with a single FinFET,
and both the fins and substrate are modeled as single-crystalline
silicon.
The geometry and material parameters used throughout this work are
summarized in Table~\ref{tab:problem_setup}.

Heat generation is localized to the fins, with each fin assigned a
prescribed ON or OFF heating state.
The resulting array therefore represents spatially heterogeneous
thermal loading over a common substrate.
At the fin scale, the characteristic dimensions become comparable to
the phonon mean free path and heat transport can depart from the
diffusive regime.
Transport within each fin is therefore modeled using the phonon BTE.
Away from these localized regions, heat transport through the much
larger substrate is assumed to be diffusive and is modeled using
Fourier conduction.

The substrate contains no volumetric heat source.
Its lateral surfaces are adiabatic, while the bottom surface, $\Gamma_{\rm sink}$, is
maintained at $T_{\rm sink}=300~\mathrm{K}$.
Heat generated within the fins enters the substrate through the
fin--substrate interfaces, which provide the coupling boundaries
between the kinetic and continuum descriptions.

\begin{table}[!t]
\centering
\caption{
Geometric and material parameters for the 100-fin thermal problem.
}
\label{tab:problem_setup}
\begin{tabular}{lll}
\toprule
Parameter & Symbol & Value \\
\midrule

Array configuration
    & $N_r\times N_c$
    & $4\times25$ \\

Number of fins
    & $N_f$
    & $100$ \\

Substrate dimensions
    & $L_x\times L_y\times L_z$
    & $33\times58\times50~\mu\mathrm{m}^3$ \\

Fin dimensions
    & $l_x\times l_y\times l_z$
    & $5\times1.2\times6~\mu\mathrm{m}^3$ \\

Row clearance
    & $s_x$
    & $3.0~\mu\mathrm{m}$ \\

Column clearance
    & $s_y$
    & $1.0~\mu\mathrm{m}$ \\

Material
    & ---
    & single-crystalline Si \\

Substrate thermal conductivity
    & $\kappa_s$
    & $151.8~\mathrm{W/(m\cdot K)}$ \\

Heat-sink temperature
    & $T_{\rm sink}$
    & $300~\mathrm{K}$ \\

ON-state volumetric heat source
    & $\dot{q}_{\mathrm{on}}$
    & $7\times 10^{15} \mathrm{W}/\mathrm{m}^3$  \\

\bottomrule
\end{tabular}
\end{table}

\subsection{Bidirectional kinetic--continuum coupling}
\label{subsec:coupling}

The fin-scale BTE domains and the continuum substrate are coupled
through a non-overlapping Schwarz iteration.
Let $\Omega_s$ denote the substrate and
$\Omega_f^{(i)}$, $i=1,\ldots,N_f$, denote the $N_f=100$ fin domains,
sharing interfaces
\begin{equation}
    \Gamma_i
    =
    \partial\Omega_f^{(i)}
    \cap
    \partial\Omega_s,
    \qquad i=1,\ldots,N_f ,
\end{equation}
where $\partial\Omega$ denotes the boundary of a domain $\Omega$.
At iteration $k$, the substrate and the fins exchange data across each
$\Gamma_i$ through the following transmission conditions:
\begin{align}
    T_{f,i}^{(k)}
    &=
    T_s^{(k-1)}\big|_{\Gamma_i}
    &&\text{on } \Gamma_i ,
    \label{eq:dn_trace}
    \\
    \widehat{J}_{\Gamma,i}^{(k)}
    &=
    \mathbf{J}_{f,i}^{(k)}\cdot\mathbf{n}_i
    &&\text{on } \Gamma_i ,
    \label{eq:dn_flux}
    \\
    \left(-\kappa_s\nabla T_s^{(k)}\right)\cdot\mathbf{n}_i
    &=
    \omega_k\,\widehat{J}_{\Gamma,i}^{(k)}
    +(1-\omega_k)\,J_{\Gamma,i}^{(k-1)}
    &&\text{on } \Gamma_i .
    \label{eq:dn_sub}
\end{align}
Here $T_{f,i}$ and $\mathbf{J}_{f,i}$ are fin $i$'s own temperature and
heat flux, $T_s$ is the substrate temperature, $\kappa_s$ is the
substrate conductivity, and $\mathbf{n}_i$ is the outward unit normal
on $\Gamma_i$. Equation~\eqref{eq:dn_trace} passes
the interface temperature trace from the substrate to each fin, where it is imposed as a thermalizing condition on the fin's bottom face.
Equation~\eqref{eq:dn_flux} extracts the fin-resolved normal heat flux
$\widehat{J}_{\Gamma,i}$, and Eq.~\eqref{eq:dn_sub} imposes on the
substrate this flux relaxed against the previous iterate, using
Fourier's law and the Aitken factor $\omega_k$ of
Eq.~\eqref{eq:aitken_update}. We write $J_{\Gamma,i}^{(k)}$ for the
right-hand side of Eq.~\eqref{eq:dn_sub}, the flux imposed on the
substrate at iteration $k$. This is a Dirichlet--Neumann exchange:
temperature is imposed on the kinetic subproblem, while the
fin-resolved flux is imposed on the continuum subproblem.

On the substrate side, we use FEniCSx~\cite{baratta2023dolfinx} for the mixed FE formulation to discretize the temperature and flux field. On the fin side, by contrast, temperature and heat flux are not primary unknowns, they are moments of the phonon distribution function, obtained from GiftBTE, a phonon BTE solver which computes the macroscopic quantities of interest, required for coupling. 

\begin{figure}[tp]
    \centering
    \includegraphics[width=0.9\linewidth]{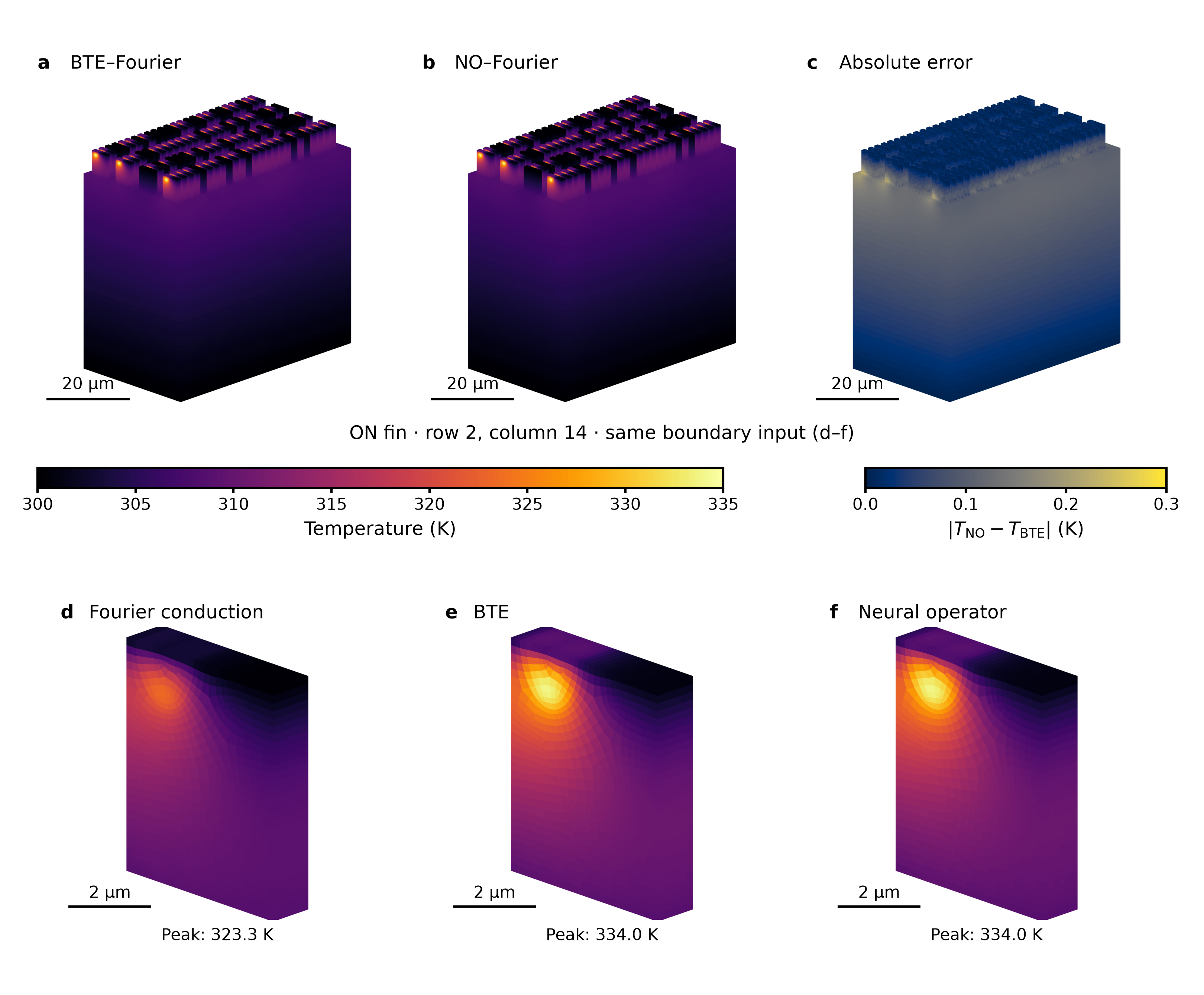}
    \caption{
    Dirichlet--Neumann coupling with Aitken-relaxed flux for an unseen
    test realization with 62 of 100 fins ON.
    BTE--Fourier and NO--Fourier agree closely over the full array
    with absolute difference under 0.3 K (c). A representative
    ON fin shows Fourier conduction (d), BTE (e) and the neural operator (f),
    all driven by the same boundary input: the neural operator matches the BTE
    peak (334.0 K) while Fourier conduction underpredicts it (323.3 K).
    }
    \label{fig:coupling_comparison_3d}
\end{figure}

\begin{table}[!t]
\centering
\caption{
Accuracy and computational cost of the full 100-fin coupled problem.
Temperature difference is the pointwise absolute error over all fin cells.
Flux difference is the pointwise $|\Delta \mathbf{J}|$ normalized by
$\max |\mathbf{J}_{\mathrm{BTE}}|$ over all fin cells.
The substrate LU factorization is computed once, reused in every iteration, and excluded from the $180\times$ speedup as a one-time cost shared across configurations. Hardware and thread counts are
given in Appendix~\ref{app:hardware}.
}
\label{tab:end_to_end}
\begin{tabular}{lcc}
\toprule
Quantity & BTE--Fourier & NO--Fourier \\
\midrule

Schwarz iterations
& 8 & 6 \\

Substrate LU factorization
& 7 min 52 s & 5 min 25 s \\

Total wall-clock
& 2 h 35 min 37 s & 6 min 14 s \\

Per iteration
& 18.4 min & 7.1 s \\

Temperature difference
& --- & 0.074 K mean, 0.290 K max \\

Flux difference
& --- & 0.074\% mean, 0.616\% max \\

Total speedup
& $1\times$ & $25.0\times$ \\

Speedup, w/o LU factorization
& $1\times$ & $180\times$ \\

Per-iteration speedup
& $1\times$ & $157\times$ \\

\bottomrule
\end{tabular}
\end{table}

To stabilize the fixed-point iteration, the flux passed to the
substrate is relaxed with an adaptive vector Aitken update, avoiding heuristics based under-relaxation
\cite{Kuttler2008DynamicRelaxation,Garbey2005AitkenSchwarz}.
Let $J_{\Gamma}^{(k-1)}$ be the concatenated flux imposed on all
interfaces in the previous substrate solve and let
$\widehat{J}_{\Gamma}^{(k)}$ be the newly computed fin flux. We form
the pooled residual
$\mathbf{r}^{(k)}=\widehat{J}_{\Gamma}^{(k)}-J_{\Gamma}^{(k-1)}$
and update the relaxation factor by
\begin{equation}
    \omega_k
    =
    -\omega_{k-1}
    \frac{
    \left\langle \mathbf{r}^{(k-1)},
    \mathbf{r}^{(k)}-\mathbf{r}^{(k-1)} \right\rangle
    }{
    \left\|\mathbf{r}^{(k)}-\mathbf{r}^{(k-1)}\right\|_2^2
    },
    \qquad
    0\le \omega_k \le 1 ,
    \label{eq:aitken_update}
\end{equation}
with $\omega_0=1$. The full
procedure, using the fixed settings $T_0$, $k_{\max}$, $\omega_0$,
$\mathrm{tol}_T$, and $\mathrm{tol}_J$, is given in
Algorithm~\ref{alg:schwarz} of Appendix~\ref{app:schwarz_algorithm}. All coupling runs use $T_0=300~\mathrm{K}$, $\omega_0=1$,
$\mathrm{tol}_T=10^{-2}~\mathrm{K}$, and $\mathrm{tol}_J=10^{-3}$.

Replacing the BTE solver with the neural surrogate leaves the rest of
the Schwarz iteration unchanged, as shown in Algorithm~\ref{alg:schwarz}
(Appendix~\ref{app:schwarz_algorithm}):
the $N_f$ independent $\mathrm{BTE}_i$ calls become a single batched
$\mathcal{G}_\theta$ query, which is evaluated directly at the FEM
interface nodes.

Figure~\ref{fig:coupling_comparison_3d} summarizes the results
obtained with this coupling strategy.

\subsection{Convergence behavior and interface consistency}
\label{subsec:interface_consistency}

\begin{figure}[tp]
    \centering
    \includegraphics[width=0.82\linewidth,keepaspectratio]{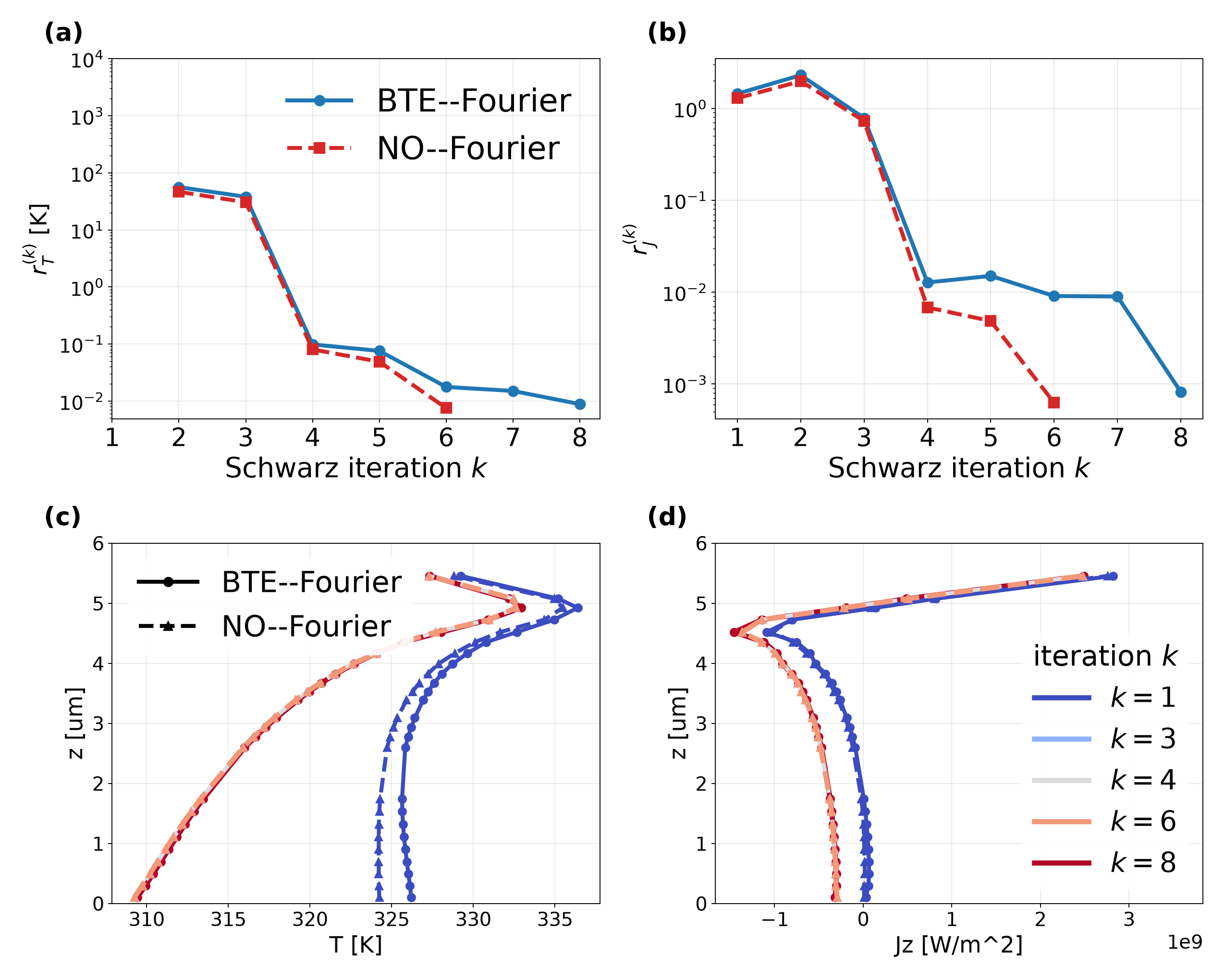}
    \caption{
    Convergence of the Aitken-relaxed Schwarz iteration for BTE--Fourier
    and NO--Fourier coupling. NO--Fourier reaches
    $\mathrm{tol}_T=10^{-2}$~K and $\mathrm{tol}_J=10^{-3}$ in six
    iterations, while BTE--Fourier requires eight iterations to meet the same tolerance condition.
    }
    \label{fig:schwarz_convergence}
\end{figure}

\begin{figure}[!t]
    \centering
    \includegraphics[width=\linewidth,height=0.70\textheight,keepaspectratio]{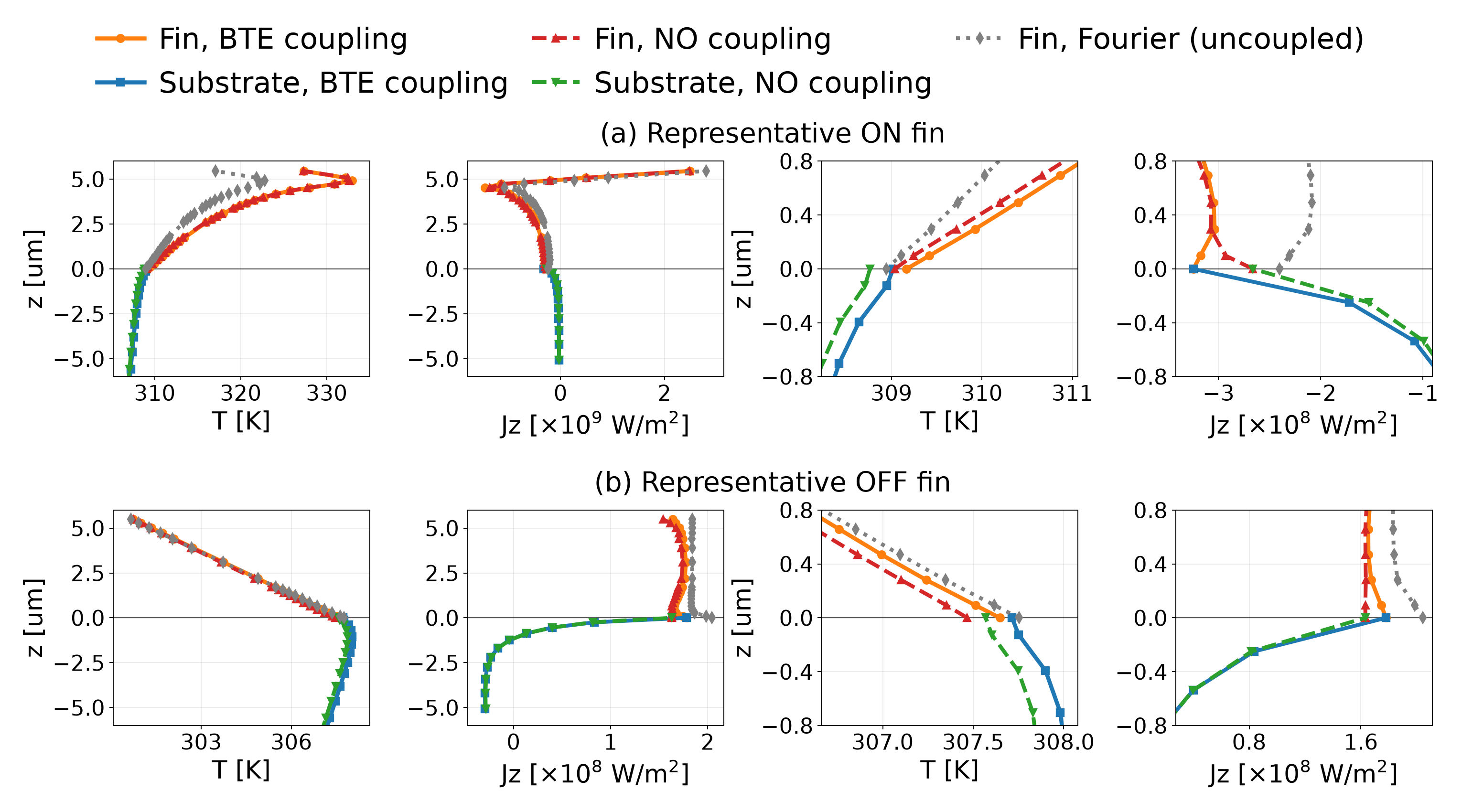}
    \caption{
    Temperature and normal heat flux continuity profiles at the interface
    for a representative ON fin (a) and OFF fin (b). BTE and NO coupling 
    agree closely and are continuous across $z=0$ in both quantities, 
    while the uncoupled Fourier fin prediction is shown to highlight 
    its departure from the true physics inside the fin.
    }
    \label{fig:interface_profiles}
\end{figure}

In Figure \ref{fig:schwarz_convergence}, we observe both frameworks show a smooth convergence in the temperature and flux residuals with Schwarz iterations, NO--Fourier converging 2 iterations faster than its BTE--Fourier counterpart for this particular realization. At convergence, we quantify the temperature mismatch and energy conservation at the interface as follows:
\begin{equation}
    \Delta T_i
    =
    \frac{1}{N_q}\sum_{q=1}^{N_q}
    \Big|T_{f,i}(\mathbf{x}_q)-T_s(\mathbf{x}_q)\Big| ,
    \qquad
    \delta Q_i
    =
    \frac{\big|Q_{f,i}-Q_{s,i}\big|}{|Q_{f,i}|}\times100 ,
    \qquad
    Q_{f,i}
    =
    \int_{\Gamma_i}\mathbf{J}_{f,i}\cdot\mathbf{n}_i\,dA ,
    \label{eq:interface_mismatch}
\end{equation}
where $\Delta T_i$ is the mean absolute mismatch over fin $i$'s
$N_q=150$ interface points $\mathbf{x}_q$, and $Q_{s,i}$ is defined
identically to $Q_{f,i}$ from the substrate side.
In the mixed-FE substrate solve, the interface flux is imposed as an
essential condition. Because the Neumann data are Aitken-relaxed during
the Schwarz iteration, the imposed flux and the raw fin flux are not
identical at a finite stopping tolerance, but they coincide in the limit of vanishing tolerance. Table~\ref{tab:interface_consistency} shows that the
remaining mean mismatch is only a few $10^{-3}\%$. The temperature
mismatch is also minimal (Fig.~\ref{fig:interface_profiles}). Besides the finite stopping tolerance, the interface temperature deviates because of the near-boundary Knudsen layer
\cite{Peraud2016}.

\begin{table}[!t]
\scriptsize
\setlength{\tabcolsep}{3pt}
\centering
\caption{
Interface consistency after convergence, evaluated independently from
the fin and substrate solutions over all fin--substrate interfaces.
}
\label{tab:interface_consistency}
\begin{tabular}{llcccccc}
\toprule
&
&
\multicolumn{3}{c}{$|\Delta T|$ (K)}
&
\multicolumn{3}{c}{$\delta Q$ (\%)} \\
\cmidrule(lr){3-5}
\cmidrule(lr){6-8}
Coupling & Fin state
& Mean & Median & Max
& Mean & Median & Max \\
\midrule

\multirow{3}{*}{BTE--Fourier}
& ON
& 0.125 & 0.124 & 0.139
& 0.007 & 0.007 & 0.010 \\
& OFF
& 0.064 & 0.064 & 0.076
& 0.013 & 0.004 & 0.311 \\
& All
& 0.101 & 0.119 & 0.139
& 0.009 & 0.007 & 0.311 \\

\midrule
\multirow{3}{*}{NO--Fourier}
& ON
& 0.152 & 0.151 & 0.176
& 0.003 & 0.002 & 0.007 \\
& OFF
& 0.086 & 0.086 & 0.101
& 0.005 & 0.004 & 0.015 \\
& All
& 0.127 & 0.143 & 0.176
& 0.003 & 0.003 & 0.015 \\

\bottomrule
\end{tabular}
\end{table}

\subsection{Overview of the fin-scale neural operator}
\label{subsec:operator_accuracy}

The pretrained neural operator, a DeepONet, substitutes the phonon BTE solver in the coupling and maps the interface temperature trace
and fin heating state to the temperature and heat flux throughout the
fin,
\begin{equation}
    \mathcal{G}_{\theta}:
    (T_\Gamma,\mathbbm{1}_{\{\dot{q}=\dot{q}_{\mathrm{on}}\}})
    \longmapsto
    (T,J_x,J_y,J_z),
\end{equation}
where $\mathbbm{1}_{\{\dot{q}=\dot{q}_{\mathrm{on}}\}}\in\{0,1\}$ is the
indicator of the fin's heating state. We compute both volumetric and interface accuracy of the operator. For each channel $c$, held-out test-set accuracy is reported as
\begin{equation}
\epsilon_c = \frac{1}{N\,\mathrm{RMS}_c}\sum_{n=1}^{N}
\left(\frac{1}{P}\sum_{p=1}^{P}\big(\hat y_c^{(n,p)}-y_c^{(n,p)}\big)^2\right)^{1/2},
\quad
\mathrm{RMS}_c = \left(\frac{1}{NP}\sum_{n=1}^{N}\sum_{p=1}^{P}\big(y_c^{(n,p)}\big)^2\right)^{1/2},
\label{eq:eps_metric}
\end{equation}
where $N$ is the number of test samples, $P$ is the number of queried points per sample ($150$ at the interface, $4692$ in the fin volume), and $y_c^{(n,p)}$ 
and $\hat y_c^{(n,p)}$ are the ground-truth and predicted values of channel $c$ at point $p$ in sample $n$.

To impose energy conservation on the operator's prediction, we construct an energy loss metric 
using first principles control volume analysis on the fin, as given in Eq. (\ref{eq:cvanalysis}).
\begin{equation}
    \int_{\partial\Omega_{\rm fin}} \widehat{\mathbf{J}}(\mathbf{x})\cdot\mathbf{n} dS = \int_{\Omega_{\rm fin}}\dot{q}_{\rm on} \, dV = Q_{\mathrm{gen}}
    \label{eq:cvanalysis}
\end{equation}
\begin{equation}
Q_{\mathrm{out}}^{(n)} = \int_{\partial\Omega_{\rm fin}} \widehat{\mathbf{J}}(\mathbf{x})\cdot\mathbf{n} dS \approx \sum_{f=1}^{6}
\sum_{i=1}^{n_{\mathrm{GL}}}\sum_{j=1}^{n_{\mathrm{GL}}}
w_{ij}^{(f)}\,
\widehat{\mathbf{J}}\big(\mathbf{x}_{ij}^{(f)}\big)\cdot\mathbf{n}_f,
\label{eq:qout}
\end{equation}
where $f$ indexes the fin boundary faces, $\mathbf{n}_f$ is the outward
normal, and $\{\mathbf{x}_{ij}^{(f)},w_{ij}^{(f)}\}$ are the nodes and weights of a
tensor-product Gauss--Legendre rule with $n_{\mathrm{GL}}\times n_{\mathrm{GL}}$
nodes on face $f$, whose weights sum to the face area $A_f$.
The heat generated in an ON fin is
$Q_{\mathrm{gen}}=\dot{q}_{\mathrm{on}}V_{\mathrm{h}}=1.972\times10^{-2}$~W, where $V_{\mathrm{h}}$ is the
volume of the fin in which the heat source is active; an OFF fin generates no heat.
The energy residual is
\begin{equation}
\epsilon_Q = \frac{1}{N\,Q_{\mathrm{gen}}}\sum_{n=1}^{N}
\left|Q_{\mathrm{out}}^{(n)}
- \mathbbm{1}_{\{\dot{q}=\dot{q}_{\mathrm{on}}\}}^{(n)}\,Q_{\mathrm{gen}}\right|,
\label{eq:eps_energy}
\end{equation}
which can be augmented into the loss function, training the operator with a physics-informed regularization.
The reported $\epsilon_Q$ uses $n_{\mathrm{GL}}=48$. During training, the face integral is instead estimated
from randomly sampled points to avoid overfitting at the GL nodes (Section~\ref{subsec4_4}).

\begin{table}[b]
\centering
\caption{
Loss configuration and held-out accuracy of the fin-scale neural
operator. Bulk errors are reported for $T$ and $\mathbf{J}$, interface
errors for $T$ and $J_z$, and $\epsilon_Q$ measures global energy
balance.
}
\label{tab:operator_ablation}
\scriptsize
\setlength{\tabcolsep}{2pt}
\begin{tabular}{lccccccccccc}
\toprule
Model
& $w_T$
& $w_J$
& $w_\Gamma$
& $w_E$
& $\epsilon_T$
& $\epsilon_{J_x}$
& $\epsilon_{J_y}$
& $\epsilon_{J_z}$
& $\epsilon_{T,\Gamma}$
& $\epsilon_{J_z,\Gamma}$
& $\epsilon_Q$ \\
\midrule

$w_\Gamma=0$ (baseline)
& 1 & (1,1,1) & 0 & 0
& 3.6e-5 & 2.1e-3 & 1.7e-2 & 2.7e-3
& 1.7e-4 & 1.2e-1 & 3.8e-2 \\

$w_\Gamma=1$ (interface loss)
& 1 & (1,1,1) & 1 & 0
& 5.3e-5 & 2.3e-3 & 1.7e-2 & 8.5e-3
& 1.1e-5 & 1.1e-2 & 2.5e-2 \\

$w_E=1$ (energy loss)
& 1 & (1,1,1) & 0 & 1
& 3.7e-5 & 2.4e-3 & 1.6e-2 & 3.2e-3
& 1.1e-4 & 9.4e-2 & 5.3e-4 \\

$w_\Gamma=1,w_E=1$ (combined)
& 1 & (1,1,1) & 1 & 1
& 5.4e-5 & 2.5e-3 & 1.7e-2 & 9.2e-3
& 1.1e-5 & 1.1e-2 & 6.0e-4 \\

\bottomrule
\end{tabular}
\end{table}

\begin{figure}[p]
    \centering
    \includegraphics[width=\linewidth,height=0.88\textheight,keepaspectratio]{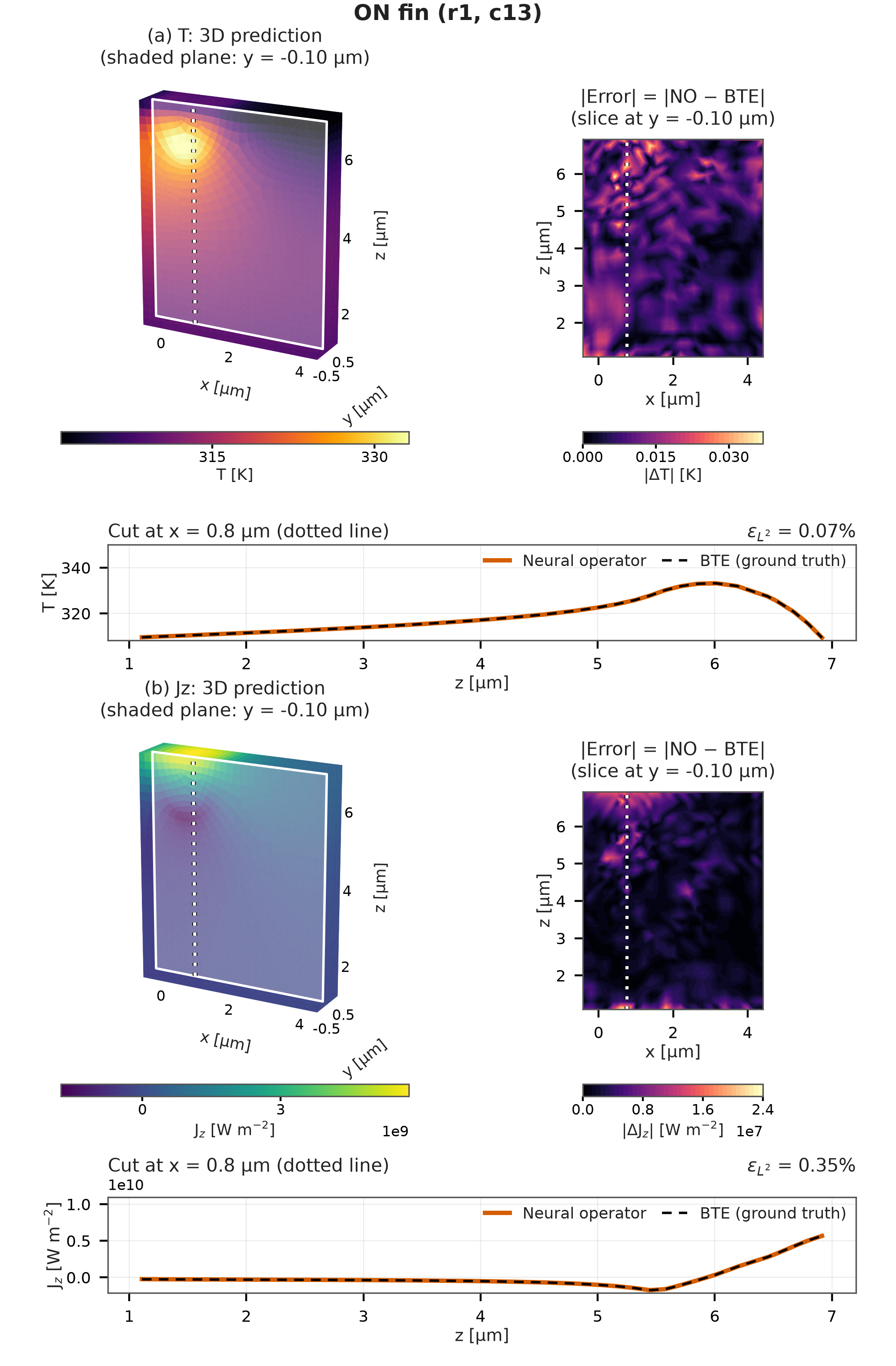}
    \caption{
    Neural operator versus BTE for an ON fin of the unseen test realization of Fig.~\ref{fig:coupling_comparison_3d}.
    (a)~$T$ and (b)~$J_z$: 3D prediction with the slice plane, absolute error on 
    that plane, and a line cut along $z$ (dotted line). The error is the absolute
    difference from the converged BTE--Fourier coupling solution. We pass the same input to the BTE solver and the neural operator to isolate the error in operator learning.
    }
    \label{fig:operator_predictions}
\end{figure}

\begin{figure}[p]
    \centering
    \includegraphics[width=\linewidth,height=0.88\textheight,keepaspectratio]{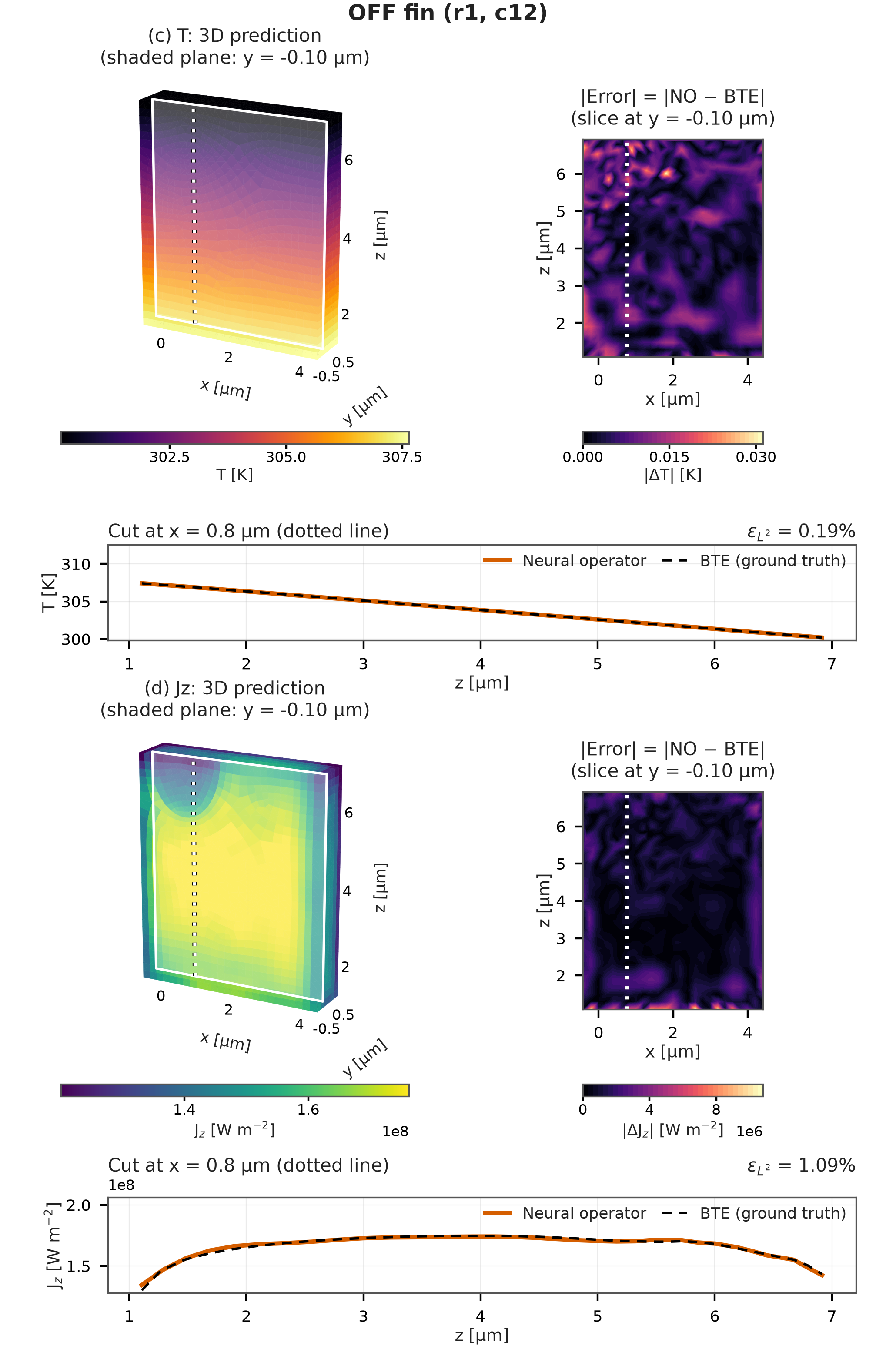}
    \caption{
    Similar analysis as Fig.~\ref{fig:operator_predictions}, for the neighboring OFF fin (row~1,
    column~12) of the same unseen realization, again with the converged BTE--Fourier interface trace as input.
    The same pretrained neural operator is used for both ON and OFF fins, with the
    heating state supplied as an input.
    }
    \label{fig:operator_predictions_slice}
\end{figure}

Table~\ref{tab:operator_ablation} presents the ablation with and without the interface loss and energy loss in the 
operator training. The baseline here refers to the purely data driven variant with error minimization over the entire fin domain.
The energy loss reduces the energy residual by about two orders of magnitude relative to the baseline while keeping the bulk field errors low. The interface loss alone leaves a large energy residual, and combining the two increases the bulk $J_z$ error. We therefore adopt the energy-loss variant, which prioritizes field accuracy and energy conservation. Its larger interface error does not impair the coupling (Table~\ref{tab:end_to_end}). Details of model architecture and training can be found in Section~\ref{sec4}. Figures~\ref{fig:operator_predictions} and~\ref{fig:operator_predictions_slice} compare the operator with the BTE solution for an ON and an OFF fin of an unseen realization.

\subsection{Generalization across switching configurations}
\label{subsec:realization_variation}

\begin{figure}[!p]
    \centering
    \includegraphics[width=\linewidth,height=0.85\textheight,keepaspectratio]{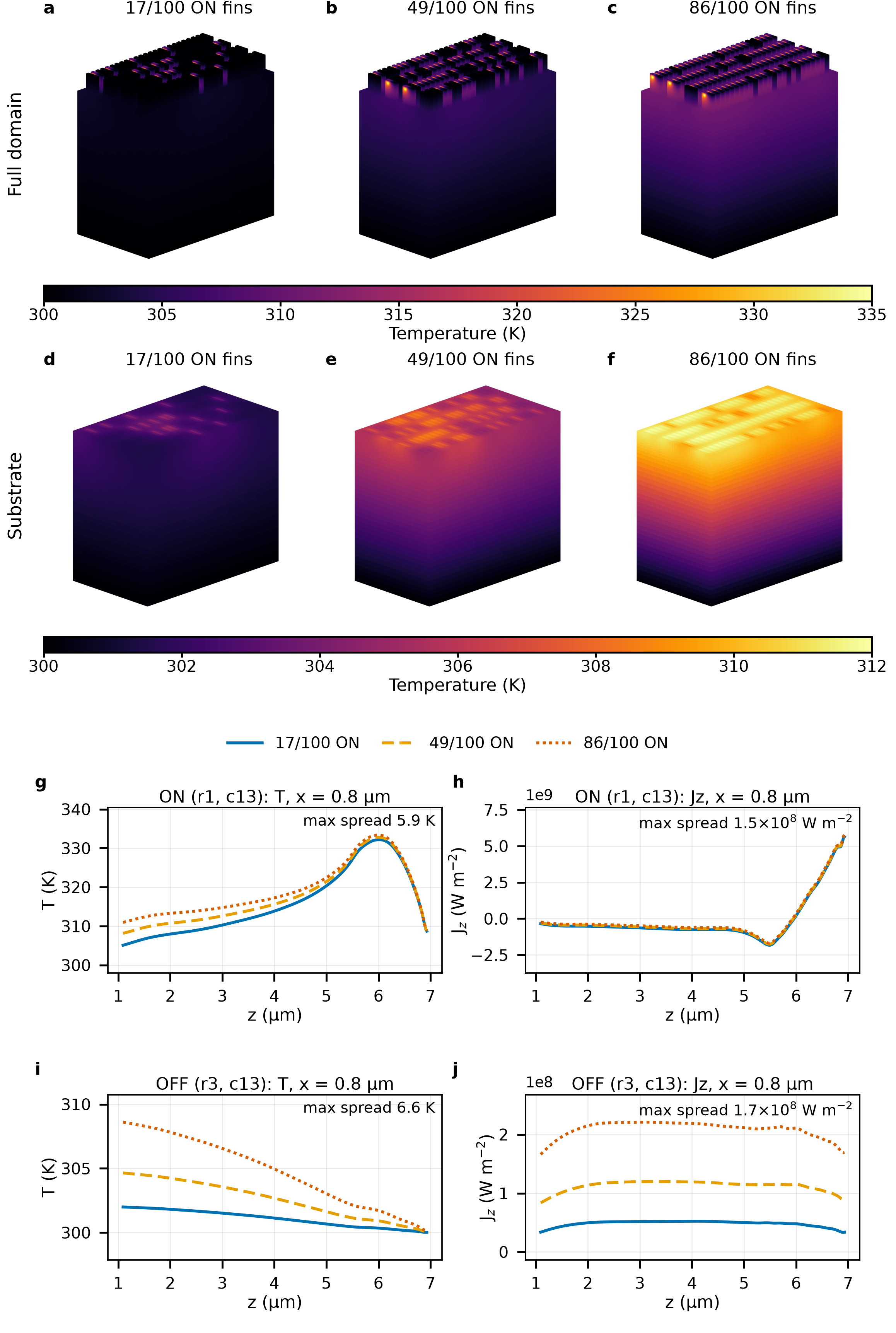}
    \caption{
    Temperature for three unseen test realizations with 17, 49 and 86 of 100 fins ON, from NO--Fourier coupling, highlighting the variation in temperature and flux across different switching configurations.
    }
    \label{fig:realization_variation}
\end{figure}

We test the full 100-fin NO--Fourier coupling on three held-out test realizations,
none of which was used in training or validation, with 17, 49 and 86 of 100 fins ON. The coupling is the non-overlapping
Dirichlet--Neumann Schwarz iteration of Section~\ref{subsec:coupling} with
Aitken relaxation, and the fins are represented by the energy-loss variant of
the operator ($w_E=1$). The coupling is stable in all three cases and converges
within 7 iterations. Fig.~\ref{fig:realization_variation} compares the
resulting temperature and flux. The same energy-loss operator, trained only on
fin-scale BTE data, therefore reproduces stable and physically consistent
couplings for heating patterns it has never seen. A direct BTE--Fourier reference is omitted for these cases because of its cost; accuracy is established on the unseen 62-ON realization (Table~\ref{tab:end_to_end}). As the number of ON fins increases from 17 to 49 and 86, the peak substrate temperature rises from $304.9$ to $308.6$ and $312.0$ K, respectively (Fig.~\ref{fig:realization_variation}d--f). The peak temperature within an ON fin also increases, though more modestly, from $333.3$ to $334.5$ K, since it is governed primarily by the local heat source. The influence of neighboring fins becomes more pronounced towards the fin--substrate interface, as seen in the temperature profiles in Fig.~\ref{fig:realization_variation}g--j. While the profiles near the hotspot differ by only $1.2$ K, their separation increases to $5.9$ K near the interface. An OFF fin, in the absence of local heating, responds more directly to changes in the surrounding thermal field. Its interface temperature rises by $6.6$ K, accompanied by an approximately $5\times$ increase in heat flux between the 17-ON and 86-ON configurations. These variations reflect the mutual dependence established by bidirectional coupling, whereby heat dissipated by individual fins alters the substrate temperature and, in turn, the thermal response of neighboring fins.

\section{Discussion}\label{sec3}

This work shows that a single pretrained neural operator can replace the
repeated BTE solves required by a large collection of kinetic subdomains, while
preserving their two-way coupling with a common Fourier substrate. For the
100-fin array, the coupled solution agrees closely with the reference
BTE--Fourier calculation in both temperature and heat flux, and remains stable
across the heating configurations considered. Excluding the one-time
substrate LU factorization cost, the coupled calculation is approximately
$180\times$ faster than direct BTE--Fourier coupling. Generating the training data and training
the operator is a one-time cost comparable to a single full BTE--Fourier coupling. This is a significant reduction, made possible by generating the interface traces from a linearized proxy (Sec.~\ref{subsec4_3}) instead of repeated couplings. Hence, break-even is reached within just two full realization solves.

The acceleration is accompanied by stable and conservative exchange across the
BTE--Fourier interface. The mixed finite element formulation permits the normal
heat flux to be imposed directly on the substrate, while the conservation
constraint used in training encourages the surrogate to respect the energy
balance within each fin. Normal-flux continuity is consequently recovered at
convergence. Since the fin and substrate are made of the same material,
the interface is fictitious and ideally should introduce no temperature
jump. However, phonons entering the fin are assigned an equilibrium
distribution at the substrate temperature, whereas those leaving the fin
have a non-equilibrium distribution. This difference produces a Knudsen
boundary layer and a small temperature slip, observed in the BTE data and naturally inherited by the trained neural operator. Adjusting the imposed BTE boundary temperature to compensate for this slip
could reduce the interface mismatch, but this correction is not investigated
here.

The present study is confined to micron-scale dimensions, for which the
division between kinetic transport in the fins and diffusive transport in the
substrate is reasonably distinct. This separation becomes less tenable at nanometer scales, where the kinetic region would need to extend beneath the fin into the substrate, moving the BTE--Fourier interface away from the geometric fin boundary. 
Such a decomposition would also make the coupling more sensitive as the fins will interact directly with the neighboring fins. The pretrained operator will therefore need to account for position or local neighborhood in addition to just the interface trace. Addressing this regime is a potential direction for future work.

Although demonstrated here for phonon transport, this approach could be
applied to other materials problems requiring coupled mesoscopic and
continuum descriptions. Local neural operators would represent the
mesoscopic material response, while exchanging boundary data with the
surrounding continuum solver.

\section{Methods}\label{sec4}

\subsection{Fin: the phonon Boltzmann transport equation}\label{subsec4_1}

Inside each fin, phonon transport is described by the phonon BTE in the
single-mode relaxation-time approximation \cite{Bhatnagar1954} and solved
with GiftBTE \cite{Hu2023}. Following the energy-based form of the phonon
BTE \cite{Loy2013}, linearized about $T_{\rm ref}=300$\,K, the unknown is the
deviational energy density
\begin{equation}
e = \frac{\hbar\omega\,D(\omega,b)\,\bigl[f - f^{0}(T_{\rm ref})\bigr]}{4\pi},
\end{equation}
where $f$ is the phonon occupation, $f^0$ the Bose--Einstein distribution,
$\omega$ the frequency, $b$ the polarization branch, and $D$ the density of
states. For phonons travelling in direction $\hat{\mathbf{s}}$, $e$ satisfies
\begin{equation}
v_g\,\hat{\mathbf{s}}\cdot\nabla e
 = \frac{e^{0}(T_L) - e}{\tau} + \dot{q},
\label{eq:ebte}
\end{equation}
where $v_g$ and $\tau$ are the group velocity and relaxation time of the
mode, $\dot{q}$ is the volumetric heat source, and
\begin{equation}
e^{0}(\mathbf{x},\omega,b) = \frac{C(\omega,b)\,\bigl[T_L(\mathbf{x}) - T_{\rm ref}\bigr]}{4\pi}
\label{eq:e0}
\end{equation}
is the local equilibrium energy, with $C=\partial e^0/\partial T$ the mode
heat capacity and $T_L$ the lattice temperature. Energy conservation of the
scattering term fixes $T_L$,
\begin{equation}
T_L(\mathbf{x}) = T_{\rm ref}
 + \frac{\displaystyle\sum_b\int \frac{1}{\tau}\int_{4\pi} e\;d\Omega\,d\omega}
        {\displaystyle\sum_b\int \frac{C}{\tau}\,d\omega},
\label{eq:TL}
\end{equation}
and the heat flux is
\begin{equation}
\mathbf{J}(\mathbf{x}) = \sum_b\int\!\!\int_{4\pi}
  v_g\,\hat{\mathbf{s}}\;e\;d\Omega\,d\omega .
\label{eq:flux}
\end{equation}
The angular quadrature uses $N_\theta=2$ and $N_\phi=4$, and frequency
is discretized into $15$ bands from ALAMODE silicon data at $300$\,K \cite{Tadano2014}.

The bottom face of each fin is the fin--substrate interface. At each
coupling iteration, the substrate solution provides the interface temperature,
which is passed to the fin as a thermalizing boundary condition on this face:
phonons entering the fin are emitted at an equilibrium state at the
prescribed temperature. All other fin boundaries follow the GiftBTE fin case
setup \cite{Hu2023}. For dataset generation, the targets are the fin-volume
$T_L$ and $\mathbf{J}$ returned by GiftBTE for each input trace.

\subsection{Substrate: mixed finite element formulation for Fourier conduction}\label{subsec4_2}

For the substrate, we solve the steady heat conduction equation without
volumetric heat sources. Assuming Fourier's law with an isotropic thermal
conductivity $\kappa>0$, the governing equations are
\begin{equation}
\mathbf{J} = -\kappa\nabla T,
\qquad
\nabla\cdot\mathbf{J} = 0
\quad \text{in } \Omega.
\label{eq:fourier}
\end{equation}
The boundary is partitioned as
$\partial\Omega = \overline{\Gamma_{\rm sink}}\cup\overline{\Gamma_N}$, with
$T = T_{\rm sink}$ on $\Gamma_{\rm sink}$ and a prescribed normal heat flux
$\mathbf{J}\cdot\mathbf{n} = g_N$ on $\Gamma_N$ ($g_N = 0$ on
insulated surfaces). We solve \eqref{eq:fourier} in mixed form, treating the heat flux
$\mathbf{J}$ and the temperature $T$ as independent unknowns. Let
\begin{equation}
V = H(\mathrm{div};\Omega)
  = \bigl\{\boldsymbol{\tau}\in[L^2(\Omega)]^d :
    \nabla\cdot\boldsymbol{\tau}\in L^2(\Omega)\bigr\},
\qquad
W = L^2(\Omega),
\end{equation}
and define
\begin{equation}
V_g = \bigl\{\boldsymbol{\tau}\in V : \boldsymbol{\tau}\cdot\mathbf{n} = g_N
      \ \text{on } \Gamma_N\bigr\},
\qquad
V_0 = \bigl\{\boldsymbol{\tau}\in V : \boldsymbol{\tau}\cdot\mathbf{n} = 0
      \ \text{on } \Gamma_N\bigr\}.
\end{equation}
Multiplying $\kappa^{-1}\mathbf{J} + \nabla T = \mathbf{0}$ by
$\boldsymbol{\tau}\in V_0$ and integrating by parts, and testing the
conservation law against $v\in W$, gives the weak form: find
$(\mathbf{J},T)\in V_g\times W$ such that
\begin{align}
\int_\Omega \kappa^{-1}\mathbf{J}\cdot\boldsymbol{\tau}\,d\mathbf{x}
 - \int_\Omega T\,\nabla\!\cdot\!\boldsymbol{\tau}\,d\mathbf{x}
 &= -\int_{\Gamma_{\rm sink}} T_{\rm sink}\,\boldsymbol{\tau}\cdot\mathbf{n}\,ds
 &&\forall\,\boldsymbol{\tau}\in V_0, \label{eq:mixed1}\\
\int_\Omega \bigl(\nabla\!\cdot\!\mathbf{J}\bigr)\,v\,d\mathbf{x}
 &= 0 &&\forall\,v\in W. \label{eq:mixed2}
\end{align}
In this formulation the sink temperature is imposed naturally through the
right-hand side of \eqref{eq:mixed1}, whereas the flux condition on
$\Gamma_N$ is imposed essentially through $V_g$; since
$\Gamma_{\rm sink}\neq\emptyset$, the problem is well posed.

We discretize the flux with the lowest-order Raviart--Thomas space
($\mathrm{RT}_1$ in the FEniCS/Basix degree convention, $\mathrm{RT}_0$ in the
original indexing) and the temperature with piecewise constants
($\mathrm{DG}_0$), a pair that satisfies the Babu\v{s}ka--Brezzi (inf--sup) condition \cite{Babuska1973,Brezzi1974}. The resulting saddle-point
system is assembled and solved in FEniCSx (dolfinx), with the normal-flux
condition on $\Gamma_N$ imposed as a Dirichlet condition on the
$\mathrm{RT}$ degrees of freedom.

This choice of finite element space, imposes normal flux as an essential
condition on the substrate side, so the converged Schwarz iteration
strongly enforces flux continuity at the shared interface. Substrate mesh details are given in Appendix~\ref{app:substrate_mesh}.

\subsection{Training-data generation}\label{subsec4_3}

Training the operator requires interface temperatures that resemble those seen
during coupling. Collecting them from full BTE--Fourier runs is too expensive,
so we estimate them instead. Fin solves at 300\,K and 305\,K give the interface
flux of an ON and an OFF fin and its linear sensitivity to interface
temperature. For each switching pattern, a single self-consistent interface
temperature $T^*$ follows analytically from this linear model, and the fluxes
evaluated at $T^*$ are imposed on the substrate, giving a spatially varying
temperature trace at every fin. Over $200$ switching patterns this yields
$20{,}000$ candidate traces, from which $k$-means++ selects $800$ diverse
samples ($560$ for training, $120$ for validation and $120$ for testing, split
by realization). Only these samples are solved with GiftBTE to obtain the
fin-volume targets. The formulation is given in
Appendix~\ref{app:tstar_generation}, and the cost is reported in
Appendix~\ref{app:hardware}.

\subsection{Operator architecture and training}\label{subsec4_4}

The neural operator is a DeepONet \cite{Lu2021} with four output channels: the
temperature $T$ and the three flux components $J_x$, $J_y$, $J_z$. The branch
is a linear layer with input dimension $151$, namely the $150$ interface
temperatures and one scalar indicating the ON/OFF state of the fin, and maps
it to a $256$-dimensional latent vector for each output channel. The trunk is a SIREN \cite{Sitzmann2020}
evaluated at normalized cell coordinates $\mathbf{x}$,
\begin{equation}
\mathbf{h}_0=\mathbf{x},
\qquad
\mathbf{h}_{l+1} = \sin\bigl(\omega_0(W_l\mathbf{h}_l+\mathbf{b}_l)\bigr),
\end{equation}
with widths $(64,128,256)$ and $\omega_0=5$.
Training minimizes relative $L^2$ errors in the fin volume and at the
interface. For channel $c$ and query set $D$, define
\begin{equation}
\mathcal{L}_c^{D}
=
\frac{
\left(
\sum_{n=1}^{N_b}\sum_{p=1}^{P_D}
\bigl|\hat{y}_{c,D}^{(n,p)}-y_{c,D}^{(n,p)}\bigr|^2
\right)^{1/2}}
{
\left(
\sum_{n=1}^{N_b}\sum_{p=1}^{P_D}
\bigl|y_{c,D}^{(n,p)}\bigr|^2
\right)^{1/2}},
\label{eq:loss_rel_l2}
\end{equation}
where $N_b$ is the batch size. For the volumetric loss,
$D=\mathrm{vol}$ and $P_D=P_{\rm vol}$ sampled fin-volume points; for
the interface loss, $D=\Gamma$ and $P_D=P_{\Gamma}$ interface points. For
the energy loss term, let
$r^{(n)}=Q_{\mathrm{out}}^{(n)}
-\mathbbm{1}_{\{\dot{q}=\dot{q}_{\mathrm{on}}\}}^{(n)}Q_{\mathrm{gen}}$
be the energy residual of sample $n$, and let $\mathcal{B}_{\rm on}$ and
$\mathcal{B}_{\rm off}$ be the ON and OFF samples in the batch. Then
\begin{equation}
\mathcal{L}_E
=
\frac{1}{Q_{\mathrm{gen}}}
\left(
\frac{1}{|\mathcal{B}_{\rm on}|}
\sum_{n\in\mathcal{B}_{\rm on}} |r^{(n)}|^2
\right)^{1/2}
+
\left(
\frac{1}{|\mathcal{B}_{\rm off}|}
\sum_{n\in\mathcal{B}_{\rm off}} |r^{(n)}|^2
\right)^{1/2},
\label{eq:loss_energy}
\end{equation}
with $Q_{\mathrm{out}}^{(n)}$ estimated from the predicted boundary
flux by random sampling,
\begin{equation}
Q_{\mathrm{out}}^{(n)} \approx \sum_{f=1}^{6} A_f \cdot \frac{1}{P}
\sum_{p=1}^{P}
\widehat{\mathbf{J}}\big(\mathbf{x}_p^{(f)}\big)\cdot\mathbf{n}_f,
\label{eq:qout_mc}
\end{equation}
with $P=32$ points drawn uniformly on each face and redrawn at every step.
This is a Monte Carlo counterpart of Eq.~\eqref{eq:qout}, used only in training. The first term is the ON-fin residual
normalized by $Q_{\mathrm{gen}}$; the second is the OFF-fin absolute
flux residual. The full loss is
\begin{equation}
\mathcal{L} = w_T\,\mathcal{L}_T^{\rm vol}
 + w_J\!\!\sum_{c\in\{x,y,z\}}\!\!\mathcal{L}_{J_c}^{\rm vol}
 + w_\Gamma\!\!\sum_{c\in\{T,J_z\}}\!\!\mathcal{L}_c^{\Gamma}
 + w_E\,\mathcal{L}_E,
\label{eq:loss}
\end{equation}
Table~\ref{tab:operator_ablation} reports the tested values of
$w_\Gamma$ and $w_E$ and the weights used for the reported model.

Training uses Adam at an initial learning rate of $10^{-3}$, halved on
plateau (patience $300$), with a batch of $16$ samples and $1024$
trunk points resampled per step, for $10{,}000$ epochs. Training hardware and time are
reported in Appendix~\ref{app:hardware}.

\subsection*{Code and data availability}
The code and datasets supporting the findings of this study will be made publicly available in a repository upon acceptance of the manuscript. Until then, they are available from the corresponding author upon reasonable request.

\begin{appendices}

\section{Substrate mesh and interface flux transfer}
\label{app:substrate_mesh}

The substrate mesh is a graded tensor-product grid that conforms to every fin
footprint, refined around each footprint and coarsening away from the array,
with geometrically growing spacing in depth from the interface to the heat
sink. Each cell is split into six tetrahedra, and each BTE interface
quadrilateral coincides with two triangular substrate facets, giving $300$
facets per fin. The final mesh has about $5$ million tetrahedra, with about
$10$ million $\mathrm{RT}_1$ flux and $5$ million $\mathrm{DG}_0$ temperature
degrees of freedom. Since the interface is conforming, each triangular facet
lies within one BTE quadrilateral and receives its flux density, which
preserves the integrated heat transfer across the interface.

\section{Schwarz coupling algorithm}
\label{app:schwarz_algorithm}

Algorithm~\ref{alg:schwarz} lists the full two-way Schwarz iteration of
Section~\ref{subsec:coupling} for the $N_f=100$ fin array. Here
$\mathcal{I}$ interpolates using the substrate finite element basis,
while $\mathcal{P}$ linearly extrapolates the BTE flux from the cell
centroids to the interface face and assigns it to the substrate facets.

\begin{algorithm}[!t]
\caption{Two-way Schwarz coupling for the $N_f=100$ fin array}
\label{alg:schwarz}
\begin{algorithmic}[1]
\Require $T_0,\,k_{\max},\,\omega_0,\,\mathrm{tol}_T,\,\mathrm{tol}_J$
\State $T_{\Gamma,i}^{(0)} \to T_0$, \quad $i=1,\ldots,N_f$ \Comment{initial interface temperature}
\State $(T_{f,i}^{(0)},\mathbf{J}_{f,i}^{(0)}) \to \mathrm{BTE}_i(T_{\Gamma,i}^{(0)},\dot{q}_i)$
\State $J_{\Gamma,i}^{(0)} \to \mathcal{P}_{f\rightarrow s}^{(i)}\big[\mathbf{J}_{f,i}^{(0)}\cdot\mathbf{n}_i\big]$
\State $\widehat{J}_{\Gamma,i}^{(0)} \to J_{\Gamma,i}^{(0)}$
\State $(T_s^{(0)},\mathbf{J}_s^{(0)}) \to \mathrm{FEM}\big(\{J_{\Gamma,i}^{(0)}\}_{i=1}^{N_f}\big)$
\State $\omega \to \omega_0$, \quad $\mathbf{r}_{\rm old} \to \varnothing$
\For{$k=1,\ldots,k_{\max}$}
    \State $T_{\Gamma,i}^{(k)} \to \mathcal{I}_{s\rightarrow f}^{(i)}\big[T_s^{(k-1)}|_{\Gamma_i}\big]$
        \Comment{Dirichlet data for the fin}
    \If{fin solver is BTE}
        \State $(T_{f,i}^{(k)},\mathbf{J}_{f,i}^{(k)}) \to \mathrm{BTE}_i(T_{\Gamma,i}^{(k)},\dot{q}_i)$, $\;i=1,\ldots,N_f$
        \State $\widehat{J}_{\Gamma,i}^{(k)} \to \mathcal{P}_{f\rightarrow s}^{(i)}\big[\mathbf{J}_{f,i}^{(k)}\cdot\mathbf{n}_i\big]$
    \Else \Comment{pretrained neural operator}
        \State $\widehat{T}_{f,i}^{(k)},\widehat{J}_{\Gamma,i}^{(k)} \to \mathcal{G}_\theta\big(\{T_{\Gamma,i}^{(k)},\mathbbm{1}_{\{\dot{q}_i=\dot{q}_{\mathrm{on}}\}}\}_{i=1}^{N_f}\big)$
    \EndIf
    \State $\mathbf{r}^{(k)} \to \operatorname{vec}\big(\{\widehat{J}_{\Gamma,i}^{(k)}-J_{\Gamma,i}^{(k-1)}\}_{i=1}^{N_f}\big)$
    \If{$\mathbf{r}_{\rm old}\ne\varnothing$}
        \State $\omega \to -\omega\,\langle \mathbf{r}_{\rm old},\mathbf{r}^{(k)}-\mathbf{r}_{\rm old}\rangle/\|\mathbf{r}^{(k)}-\mathbf{r}_{\rm old}\|_2^2$
        \State $\omega \to \min(1,\max(0,\omega))$
    \EndIf
    \State $J_{\Gamma,i}^{(k)} \to \omega\widehat{J}_{\Gamma,i}^{(k)}+(1-\omega)J_{\Gamma,i}^{(k-1)}$
        \Comment{Aitken-relaxed Neumann data}
    \State $(T_s^{(k)},\mathbf{J}_s^{(k)}) \to \mathrm{FEM}\big(\{J_{\Gamma,i}^{(k)}\}_{i=1}^{N_f}\big)$
    \State $r_T^{(k)} \to \max_i\|T_{\Gamma,i}^{(k)}-T_{\Gamma,i}^{(k-1)}\|_\infty$
    \State $r_J^{(k)} \to \max_i\|\widehat{J}_{\Gamma,i}^{(k)}-\widehat{J}_{\Gamma,i}^{(k-1)}\|_2 \big/ \max_j\|\widehat{J}_{\Gamma,j}^{(k-1)}\|_2$
        \Comment{worst fin sets the residuals}
    \If{$r_T^{(k)}<\mathrm{tol}_T$ \textbf{and} $r_J^{(k)}<\mathrm{tol}_J$}
        \State \textbf{break}
    \EndIf
    \State $\mathbf{r}_{\rm old} \to \mathbf{r}^{(k)}$
\EndFor
\end{algorithmic}
\end{algorithm}

\section{Data generation methodology}
\label{app:tstar_generation}

The dataset traces are generated with a cheap proxy for the coupled
fixed point. Let $T_0=300$\,K and let $j_{\rm on}^0$ and
$j_{\rm off}^0$ be the ON and OFF template normal flux traces obtained
from fin solves at $T_0$. A second pair of template solves at
$T_0+\Delta T$, with $\Delta T=5$\,K, gives
\begin{equation}
s_{\rm on/off}
=
\frac{j_{\rm on/off}(T_0+\Delta T)-j_{\rm on/off}^0}{\Delta T}.
\end{equation}
For each realization, the substrate is first solved with the
$T_0$-template fluxes. If the realization has $N_{\rm on}$ ON fins and
$N_{\rm off}$ OFF fins, the imposed heat flow and the linearized change
with interface temperature are
\begin{equation}
Q_0=N_{\rm on}A_\Gamma\sum_q j_{{\rm on},q}^0 
+
N_{\rm off}A_\Gamma\sum_q j_{{\rm off},q}^0,
\qquad
\beta=
N_{\rm on}A_\Gamma\sum_q s_{{\rm on},q}
+
N_{\rm off}A_\Gamma\sum_q s_{{\rm off},q},
\end{equation}
where $A_\Gamma$ is the area of one interface quadrilateral. The mean
interface rise from this substrate solve, $\Delta\bar{T}_0$, gives a
scalar substrate thermal resistance
\begin{equation}
g=\frac{\Delta\bar{T}_0}{Q_0}.
\end{equation}
The template heat entering the substrate is then approximated as
\begin{equation}
Q(T)=Q_0+\beta(T-T_0).
\end{equation}
The proxy temperature is the scalar fixed point
$T^*=T_0+gQ(T^*)$, or
\begin{equation}
T^* = T_0 + \frac{gQ_0}{1-g\beta}.
\label{eq:tstar_proxy}
\end{equation}
The ON and OFF template fluxes are corrected to this temperature,
\begin{equation}
j_{\rm on/off}^*
=j_{\rm on/off}^0+(T^*-T_0)s_{\rm on/off},
\end{equation}
and the substrate is solved again with $j_{\rm on}^*$ on ON fins and
$j_{\rm off}^*$ on OFF fins. The resulting $150$-point temperature
trace at each fin footprint is a candidate boundary trace.

\section{Compute resources and wall-clock time}
\label{app:hardware}

All reported times are wall-clock times. CPU-bound tasks run on AMD EPYC 9684X
(96-core) nodes, and training uses one NVIDIA L40S GPU with four cores of an
AMD EPYC 9554 host and $32$\,GB of memory. Training-data generation takes
$2$\,h $38$\,min in total, including $800$ GiftBTE solves with four MPI ranks
each and $16$ running concurrently. The substrate is LU-factorized once
(direct solver, MUMPS \cite{amestoy2001mumps}) and reused across substrate solves. Training for $10{,}000$ epochs
takes about $35$\,min. In the coupled runs of Table~\ref{tab:end_to_end}, the
BTE--Fourier fin solves use four MPI ranks per case with $14$ cases at a time,
and the substrate and inference use $8$ threads (BTE--Fourier) and $16$ threads
(NO--Fourier). The neural operator runs on CPU only, as a single batched query
for all $100$ fins.

\end{appendices}

\backmatter

\bmhead{Acknowledgements}
This material is based upon work supported by the U.S. Department of Energy, Office of Science, Office of Advanced Scientific Computing Research (ASCR) under Award Number DE-SC0023191. Support was also provided by Pacific Northwest National Laboratory (operated by Battelle Memorial Institute for the U.S. Department of Energy under Prime Contract DE-AC05-76RL01830) through Subcontract No. 814377, Subcontract No. 831126, and Joint Appointment Agreement No. 801034. R.C. gratefully acknowledges the support of the Paul Sorensen Fellowship, offered by the Brown University School of Engineering. The authors acknowledge the computational resources and services at the Center for Computation and Visualization (CCV), Brown University.

This report was prepared as an account of work sponsored by an agency of the United States Government. Neither the United States Government nor any agency thereof, nor any of their employees, makes any warranty, express or implied, or assumes any legal liability or responsibility for the accuracy, completeness, or usefulness of any information, apparatus, product, or process disclosed, or represents that its use would not infringe privately owned rights. Reference herein to any specific commercial product, process, or service by trade name, trademark, manufacturer, or otherwise does not necessarily constitute or imply its endorsement, recommendation, or favoring by the United States Government or any agency thereof. The views and opinions of authors expressed herein do not necessarily state or reflect those of the United States Government or any agency thereof.

\bmhead{Competing interests}
The authors declare no competing interests.

\bmhead{Contributions}
R.C.: Conceptualization, Methodology, Software, Formal analysis, Investigation, Data curation, Writing -- original draft.
K.S.: Formal analysis, Validation, Writing -- review \& editing, Supervision.
G.E.K.: Conceptualization, Formal analysis, Validation, Resources, Writing -- review \& editing, Supervision, Project administration, Funding acquisition.

\bibliography{main}

\end{document}